\documentclass{PoS}
\usepackage{amsmath}
\usepackage{mathrsfs}

\newcommand{\dHybridR}{{\tt dHybridR}}

\title{Hybrid Simulations of the Resonant and Non-Resonant Cosmic Ray Streaming Instability}

\ShortTitle{CR Streaming: Hybrid Simulations}

\author{\speaker{Colby Haggerty}\\
        Department of Astronomy and Astrophysics, University of Chicago,\\
        5640 S Ellis Ave, Chicago, IL 60637, USA\\
        E-mail: \email{chaggerty@uchicago.edu}}

\author{Damiano Caprioli\\
        Department of Astronomy and Astrophysics, University of Chicago,\\
        5640 S Ellis Ave, Chicago, IL 60637, USA}

\author{Ellen Zweibel\\
        Department of Astronomy, University of Wisconsin-Madison,\\
        475 North Charter Street, Madison, WI 53706, US}
        
\abstract{Using hybrid simulations (kinetic ions--fluid electrons), we test the linear theory predictions of the cosmic ray (CR) streaming instability. We consider two types of CR distribution functions: a ``hot'' distribution where CRs are represented by a drifting power law in momentum and an anisotropic ``beam'' of monochromatic particles . 
Additionally, for each CR distribution we scan over different CR densities to transition from triggering the resonant to the non-resonant (Bell) streaming instability. 
We determine the growth rates of these instabilities in simulations by fitting an exponential curve during the linear stage, and we show that they agree well with the theoretical predictions as a function of wave number agree. 
We also examine the magnetic helicity as a function of time and wave number, finding a general good agreement with the predictions, as well as some unexpected non-linear features to the instability development.}

\FullConference{36th International Cosmic Ray Conference -ICRC2019-\\
		July 24th - August 1st, 2019\\
		Madison, WI, U.S.A.}

\begin{document}
\section{Introduction}
High energetic, low number density cosmic rays (CRs) are a dynamically important component of almost every diffuse astrophysical plasma system. In the interstellar medium they make up as much as 1/3 the total energy budget.
The primary source of galactic CRs (CRs with energies less than $10^{15}$eV) is believed to be the collisionless shock waves associated with supernova remnants, through a diffusive shock acceleration mechanism (DSA) \cite{bell78a}. 
During DSA, CRs amplify magnetic fields that positively feed back on the CR acceleration \cite{bell04, caprioli+14a,caprioli+14b,caprioli+14c}. This field amplification is associated with with an instability that occurs when a CR distribution travels with a bulk drift velocity relative to the background thermal plasma, and is referred to as a \emph{streaming instability}.

The nature of such an instability depends on the physical parameters of the background plasma, the magnitude/orientation of the magnetic field, as well as the CR distributions (the momentum distribution, the number density, the anisotropy of the distribution function and the bulk flow relative to the background);
 depending on whether the excited waves are resonant ---in polarization and wavelength--- with the CR populations, we talk of \emph{resonant} and \emph{non-resonant} (or Bell) instabilities \cite{zweibel03, bell04, amato+09, zweibel13}. 

The resonant instability is generated through a gyro-resonant interaction between the magnetic fields and the CRs, while the non-resonant instability is generated by the electron current that accompanies the drifting CRs to insure quasi-neutrality. 
The boundary between these two regimes can be characterized by 
\begin{equation}\label{eq:sigma}
    \bar{\sigma} \equiv \frac{n_{cr}}{n_i}\frac{v_D p_{cr}}{m_i v_A^2} \simeq \frac{1}{2} \frac{\epsilon_{cr}}{\epsilon_B}\frac{v_D}{c},
\end{equation}
where  $n_{cr}$ and $n_i$ are the number density of the background and CRs respectively, $v_{A}$ is the Alfv\'en speed based on the background field and density, $v_D$ is the relative velocity between the two populations, $m_i$ is the ion mass, and $c$ the speed of light.
$\epsilon_{cr}$ and $\epsilon_{B}$ are the energy densities in CRs and magnetic fields.
This quantity is also closely linked to the current carried by the CRs ($J_{cr} \sim  e n_{cr}v_D$). 
When $\bar{\sigma} \gg 1$, the growth rate of the non-resonant instability is larger than the resonant growth rate; 
when $\bar{\sigma} \ll 1$, both branches grow at the same rate for quasi-isotropic CR distribution functions  \cite{amato+09}. 
Also see \cite{caprioli15p, zweibel17} for reviews of the theory of these instabilities.

We use the newly-developed hybrid code, \dHybridR~\cite{haggerty+19a}, which includes the relativistic ion dynamics, to study the linear growth of the resonant and non-resonant CR streaming instability for a collection of different CR distributions. 
We compare our results with the classical theory of CR streaming theory, which is usually worked out for quasi-isotropic, slowly drifting CR distributions, and put forward and validate a novel prediction for a beam-like CR distribution.
The latter is especially relevant near CR sources, where high-energy CRs are trying to escape the system \cite{caprioli+09b}.

\section{Linear Theory}\label{disp}
For brevity, we omit an extended review of the CR streaming instabilities and present only the functional form of the CR distributions that we consider.
The first CR distribution to be considered will be referred to as the \emph{hot} setup and is a power-law distribution with minimum momentum $p_{min}$ and a slope $\alpha$ such that:
\begin{equation}\label{eq:phi}
   f_{cr}(p) \equiv 4\pi n_{cr}\phi(p); \qquad \phi(p)\equiv  \frac{(3 - \alpha)}{p_{min}^{3-\alpha}}
\begin{cases}
    p^{-\alpha},    & \text{if } p\geq p_{min}\\
    0,              & \text{otherwise}
\end{cases}
\end{equation}
where $\alpha > 3$ and $p_{min} \gg m_iv_A$. 
Such a distribution drifts with respect to the background with velocity $v_D$.
The second distribution will be referred to as the \emph{beam} setup and is characterized by an average momentum $p_b$, a spread in momentum $\Delta p$, and a distribution in pitch angles $g(\mu)$, with $\mu\equiv\cos(\theta)$ and $\theta$ the angle between {\bf p} and {\bf B}.
In formulas,
$f_{cr}(p,\mu) \equiv 4\pi n_{cr}F(p)g(\mu)$ and
\begin{align}
    F(p) \equiv  \frac{A}{\Delta p^3} \label{eq:F}
    \exp\left\{-\frac{1}{2}\frac{(p - p_b)^2}{\Delta p^2}\right\}
    \\
    g(\mu) \equiv  
\begin{cases}
    \frac{1 + m\mu}{1 + m/2},    & \text{if } \mu \ge 0 \\
    0,                           & \text{otherwise.}\label{eq:g}
\end{cases}
\end{align}
Physically speaking, we expect hot distributions where CRs are effectively scattering (diffusing), as in a shock precursor or in the Galaxy, and beam distributions far upstream of shocks and where high-energy particles escape almost unperturbed \cite{caprioli+14b}. 

The dispersion relations for the non-resonant instability growth rate are  independent of the CR distribution (only the total CR current matters) and are well documented in the literature \cite{bell04};
the theory of the resonant instability for a drifting power law with $\alpha=4$ can be found in \cite{zweibel03} and \cite{amato+09}. 
The theoretical expectation for the the beam distribution has never been worked out before and will be presented in detail in a forthcoming paper \cite{hzc19}.

\section{Hybrid Simulations}
To test the linear predictions for both the hot and beam setups, we have performed hybrid plasma simulations (kinetic ions with massless charge-neutralizing fluid electrons) with the code \dHybridR~\cite{haggerty+19a}.
\dHybridR~is the first hybrid code able to handle relativistic ions and is perfectly suited to treat energetic CRs drifting on top of a non-relativistic background.
The values in the simulations are normalized to the background plasma parameters: lengths to the ion inertial length ($d_i \equiv c/\omega_{pi}$), times to the inverse ion cyclotron time ($\Omega_{ci}^{-1}$), and speeds to the Alv\'en speed ($v_A = B/\sqrt{4\pi m_i n_i}$). 
The systems are quasi-1D ($10^4 d_i\times 5 d_i$ in real space, but retaining all three dimensions in velocity space) and periodic in all directions. 
We use two grid cells per ion inertial length and 400 time steps per inverse cyclotron time;
the speed of light is $c = 100 v_A$. 
The background magnetic field is along the longitudinal direction and the background plasma, described with 100 particles per cell, is at rest in the simulation frame with an ion thermal velocity equal to the Alfv\'en speed.
A second ion population (the CRs) is superimposed with a distribution described by either the hot or the beam setup, with four particles per cell. 
We vary the CR density over three different values for each set up ($n_{cr} = 10^{-2},\ 10^{-3}\text{ or } 10^{-4}$) to span the transition from the non-resonant to the resonant streaming instability. 
The condition for the non-resonant modes to become dominant ($\bar{\sigma}\gg1$, see Eq.~\ref{eq:sigma}) depends on CR momentum, density and bulk drift \cite{bell04, amato+09}.

Each figure in this manuscript consists of a left and right column, corresponding to the hot or beam cases, respectively. 
For the three hot simulations, the minimum power-law momentum is chosen to be $p_{min} = m_ic$, $\alpha = 4.5$, and $v_D = 10v_A$. 
In the beam setup, the center of the Gaussian in momentum space is set to $p_b = m_ic$, the width is taken to be $\Delta p = p_b/10 = m_ic/10$ and the slope of the pitch angle distribution is $m = 1$ (Eq.~\ref{eq:g}). 

\begin{figure}[tb]\centering
\includegraphics[trim=1px 0px 0px 1px, clip=true, width=.48\textwidth]{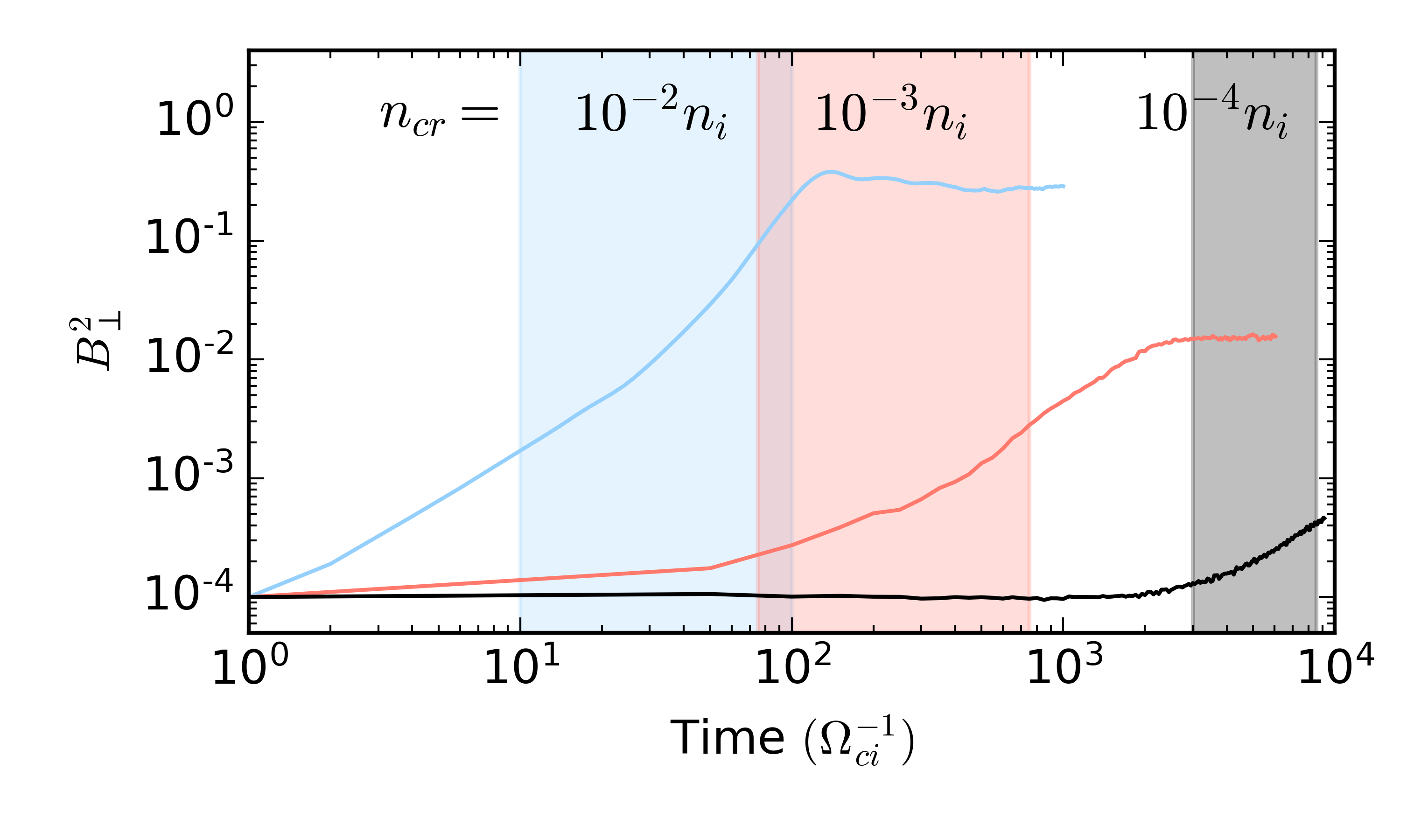}
\includegraphics[trim=1px 0px 0px 1px, clip=true, width=.48\textwidth]{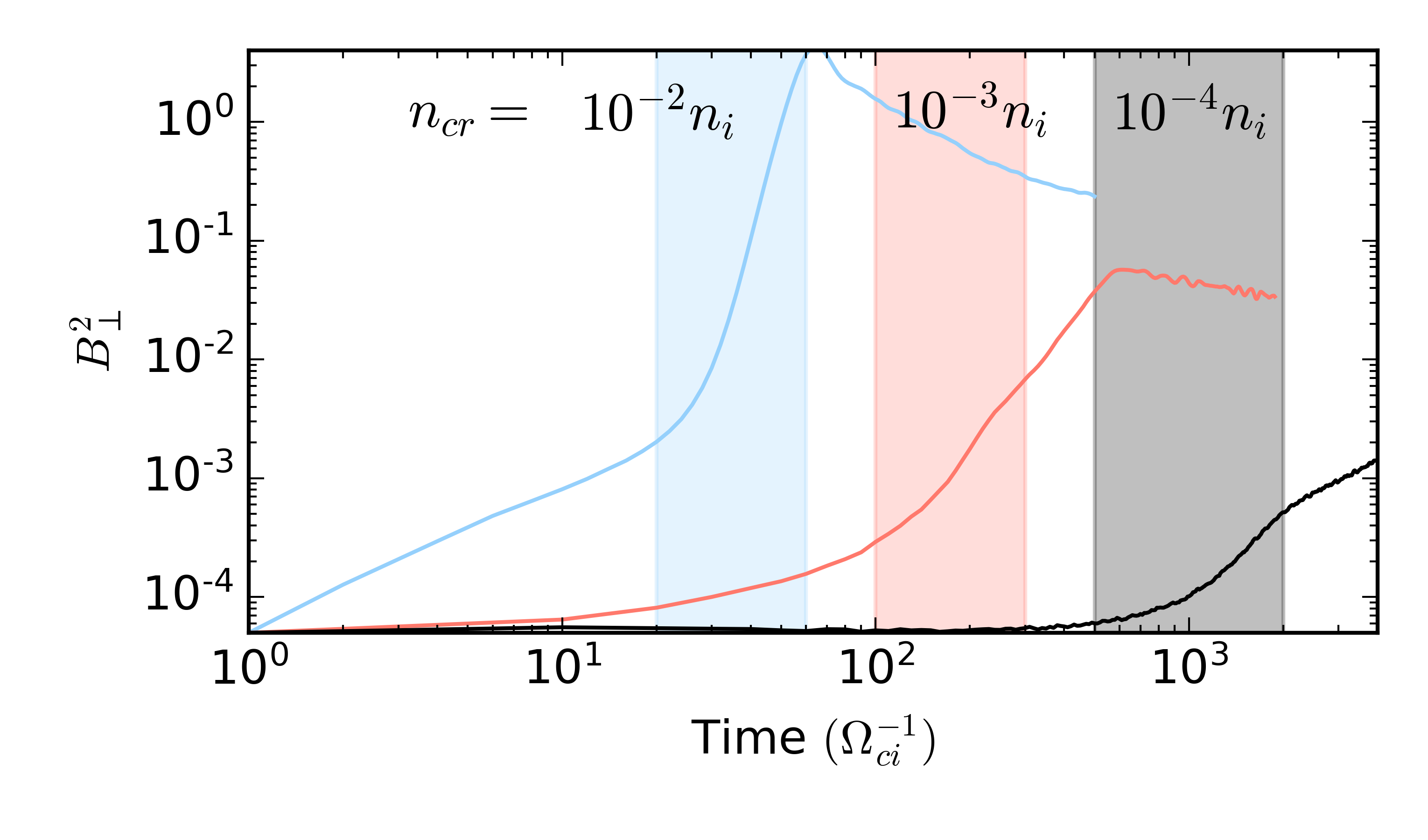}
\caption{{\small Perpendicular magnetic field squared ($B_\perp^2$) for the hot (left panel) and beam (right panel) simulations in time for three different CR densities $n_{cr}/n_i = (10^{-2},10^{-3},10^{-4}) \rightarrow$ (light blue, salmon, black). 
The shaded colored regions shows the times when the growth rates are determined.}}\label{fig:B}
\end{figure}
Each simulations is run until the perpendicular magnetic field ($B_\perp^2 = B_y^2 + B_z^2$) reaches a maximum, implying the end of a growth phase; 
this duration ranges from $10^2 $ to $ 10^4 \Omega_{ci}^{-1}$. 
The time evolution of the perpendicular magnetic energy density\footnote{Note, technically the magnetic energy density is $B^2/2$ in code units, however in this work we have neglected the factor of two, which has no impact on the analysis in this manuscript.} of the six different simulations (averaged over the entire domain) is shown in Fig.~\ref{fig:B}, where the blue, red and black lines correspond to $n_{cr} = 10^{-2},\ 10^{-3}\text{ and } 10^{-4}$, respectively. 
An instability develops in each of these simulations, and its growth rate and saturation depend on the CR distribution and density.

To make a more detailed comparison between simulations and theory, we calculate the Fourier transform (FFT) of the perpendicular magnetic fields. 
The spectral magnetic energy density is $\widetilde{B}^2_\perp(k) \equiv (|\text{FFT}[B_y(x)]|^2 + |\text{FFT}[B_z(x)]|^2)/L_x$, where the wave number $k$ is defined as $k = 2\pi/\lambda$. This normalization is chosen so that $\sum \Delta k \widetilde{B}^2_\perp = \left < B_\perp^2\right >$, the box-averaged perpendicular field.
\begin{figure}[htb]\centering
\includegraphics[trim=1px 0px 0px 1px, clip=true, width=.48\textwidth]{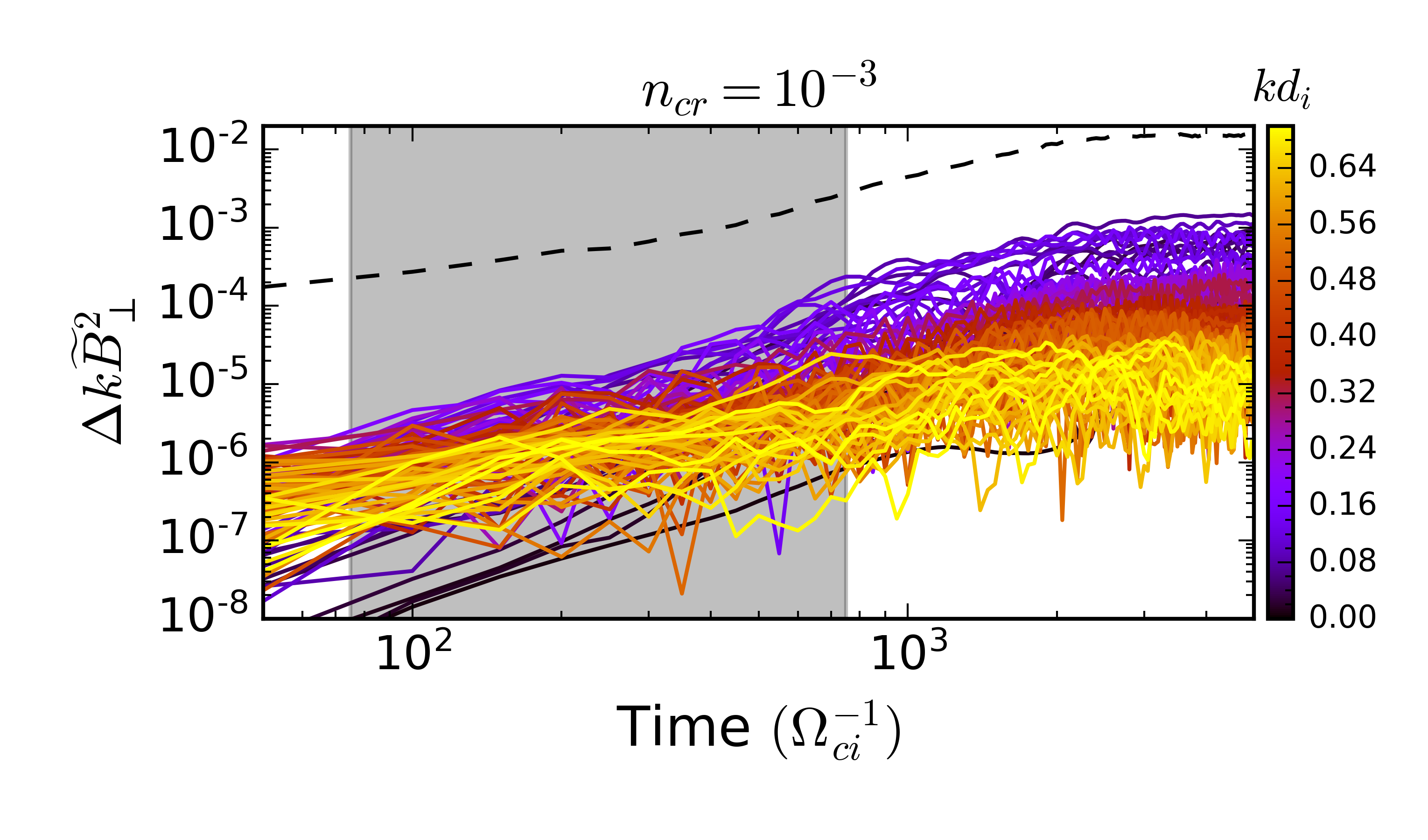}
\includegraphics[trim=1px 0px 0px 1px, clip=true, width=.48\textwidth]{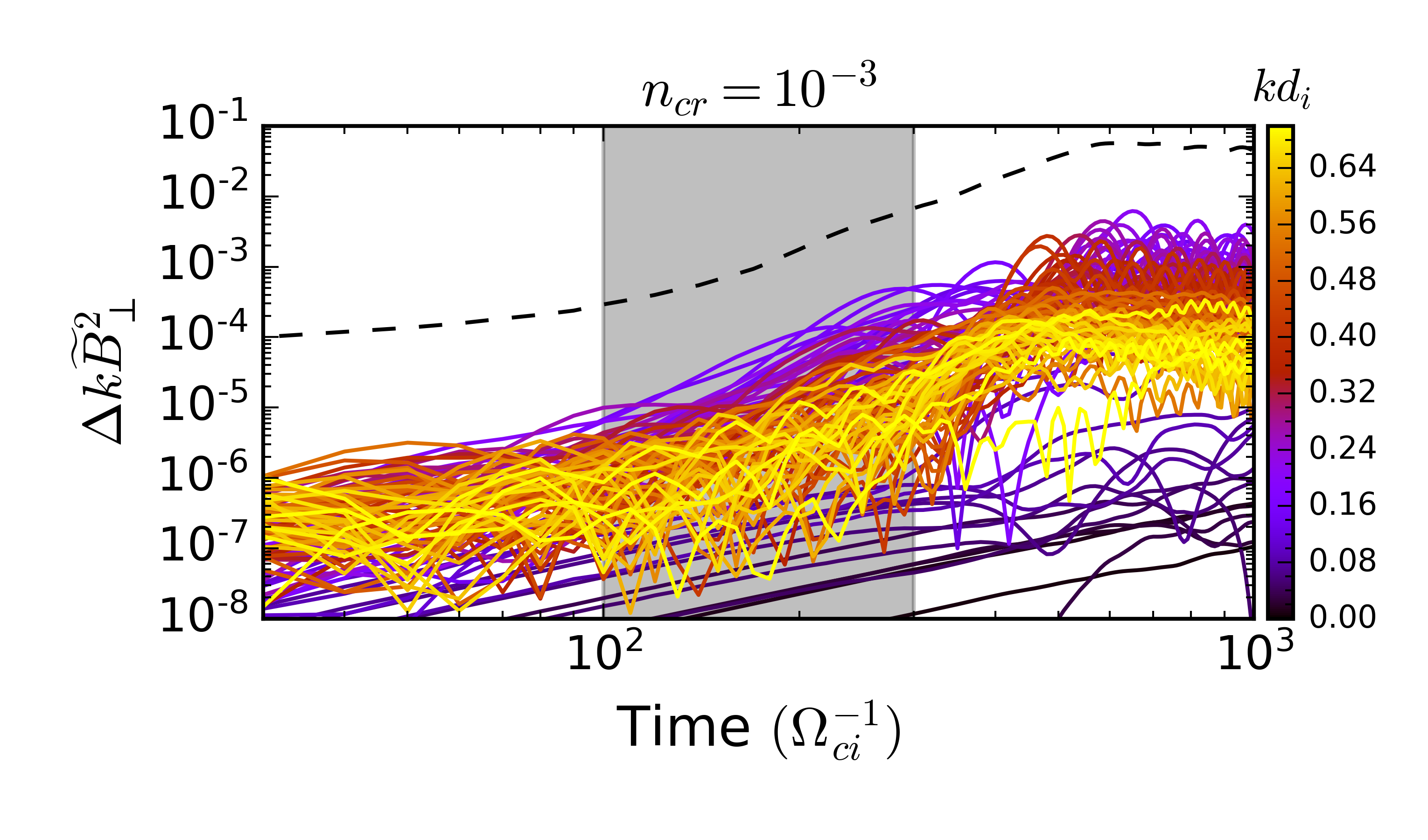}
\caption{{\small The Fourier-decomposed perpendicular magnetic field in time averaged over the entire simulation for the hot (left) and beam (right) like CR distributions for the $n_{cr} = 10^{-3}n_i$ case. The colors correspond to different wave numbers, and the black dashed line is the sum over all modes.}}\label{fig:FourDecomp}
\end{figure}
The spectral magnetic energy density (multiplied by $\Delta k$) is plotted as a function of time for the $n_{cr} = 10^{-3}n_i$ case for several different $k$ values (color coded); the sum over all $k$ values is shown by the black dashed line. 
This decomposition highlights the differences in growth rates and saturation values for different modes.

By examining the time variation of the magnetic field for a single $k$, we can quantify the growth rates of the active instabilities as a function of wave number, i.e., $\omega_I(k) \equiv {\rm Im}[\omega(k)]$; 
more precisely, we fit the spectral magnetic energy density with $\widetilde{B}_\perp^2(k) \propto e^{2\omega_I(k)t}$.
The shaded regions in Fig.~\ref{fig:B} and Fig.~\ref{fig:FourDecomp} denote which times the simulation was averaged over to determine ${\rm Im}[\omega]$.
\begin{figure}[htb]\centering
\includegraphics[trim=1px 0px 0px 1px, clip=true, width=.48\textwidth]{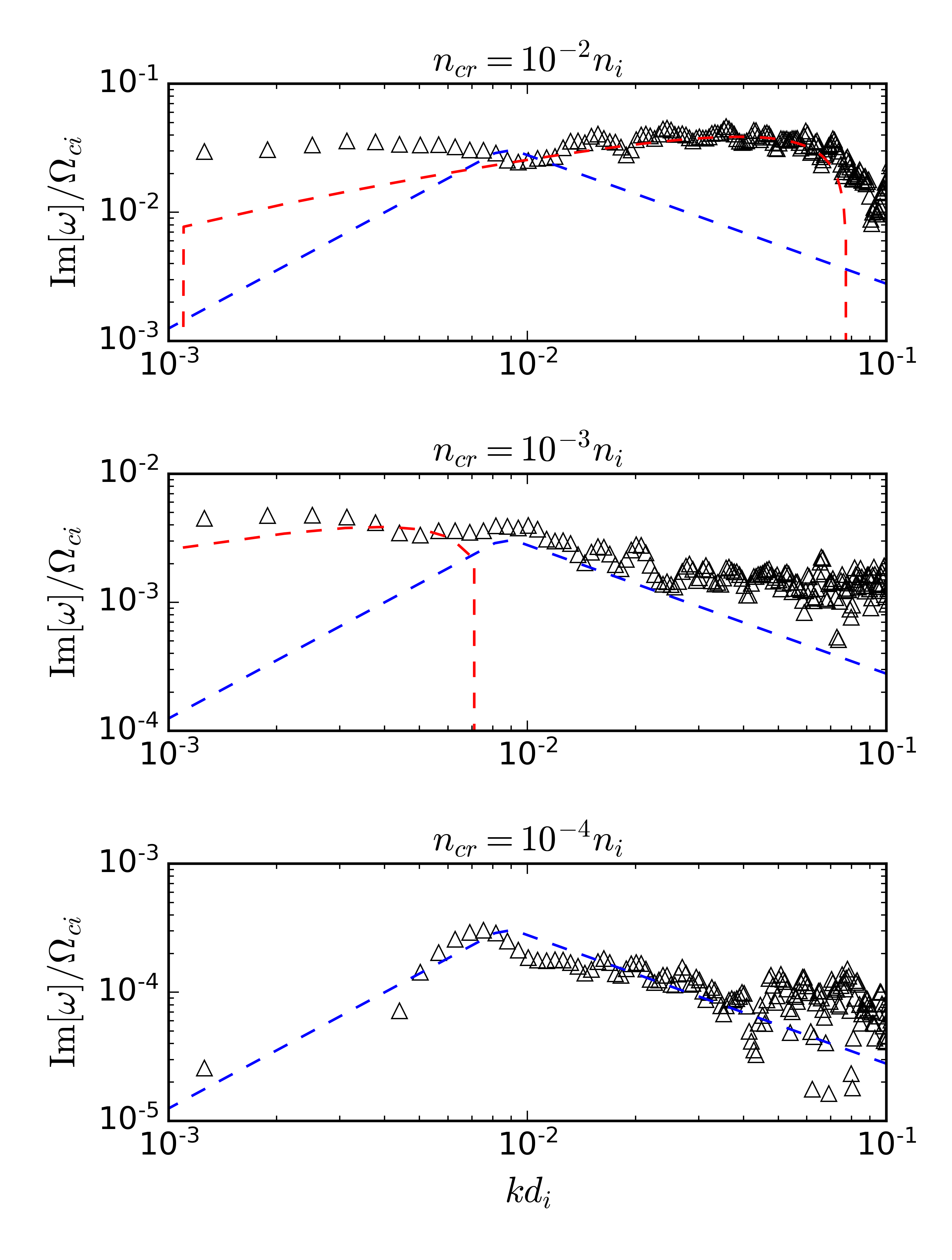}
\includegraphics[trim=1px 0px 0px 1px, clip=true, width=.48\textwidth]{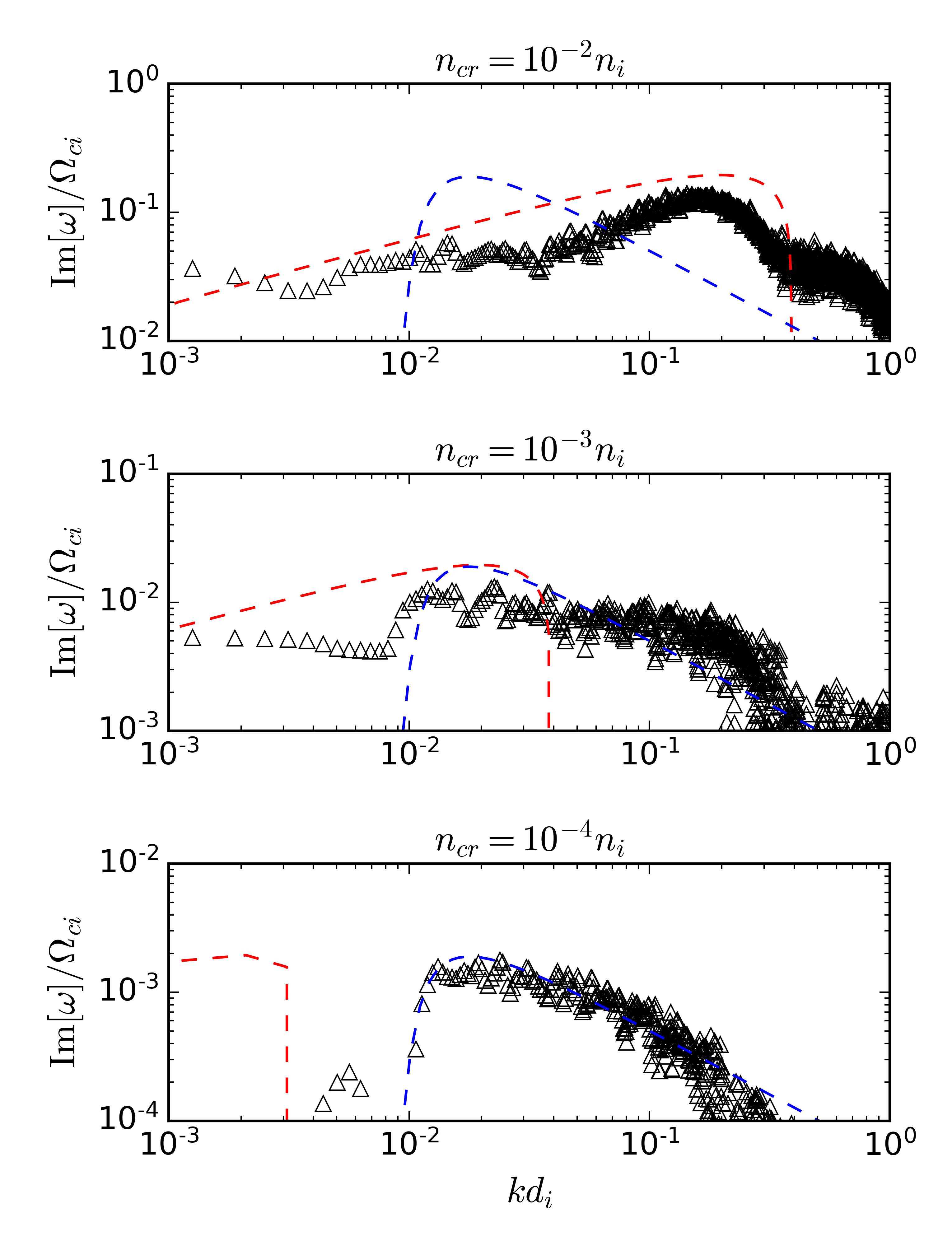}
\caption{{\small Magnetic field growth rates calculated from simulations with hot (left column) and beam (right column) distributions, for three different CR densities as in the titles. Dashed lines correspond to the linear theory prediction for the non-resonant (red) and resonant (blue) instability.}}\label{fig:growth}
\end{figure}

 The black triangles in Fig.~\ref{fig:growth} correspond to the growth rates inferred from the six runs. 
 In each of these panels, the resonant (blue dashed line) and the non-resonant (red dashed line) approximated theoretical predictions based on each simulation's CR distribution and density are shown. 
 The prediction for the non-resonant growth rate is well documented in the literature \cite{bell04,amato+09}, but the predictions for the growth rate of the resonant regime of the generalized power law and the beam case described above are original; their detailed derivations are presented in a forthcoming paper \cite{hzc19}.
 For most of the simulations the agreement is remarkable for all the wave numbers; in few cases there are some discrepancies between theory and simulations, but always within a factor of a few, very small compared to the factor of $10^4$ difference in growth rates across the different runs.

The top row of the Fig.~\ref{fig:growth} shows simulations with strong currents, well in the  non-resonant regime. 
In both the hot and beam cases, there is fairly good agreement between the red-line prediction and the simulation-observed values.
There is a notable deviation in the hot setup (upper left corner) for $kr_{g,cr} > 1$, where $r_{g,cr}\equiv p_b/m_i\Omega_{ci}$ is the gyro-radius based on the uniform magnetic field and the beam momentum. 
The growth rate of these modes appears to be significantly larger than either the non-resonant or the resonant theory predicts. The cause of this is potentially linked to the transfer of energy in $k$ space (e.g., turbulent inverse cascade) and needs to be investigated further. 
Moreover, the duration of the linear growth phase is smaller than an Alfv\'en crossing time for these modes, which could present an issue for accurately determining the growth rate.

The bottom row of Fig.~\ref{fig:growth} shows simulations expected to be in the resonant regime. 
Appropriately, the simulation agrees well with the resonant theory (blue line). In the beam set up, the linear theory predicts a sharp drop off in growth rates below $k r_{g,cr}$.

The middle row of the same figure corresponds to a parameter space that is in between the two instabilities ($n_{cr} = 10^{-3}n_i$). 
This transitional regime still shows features consistent with the linear theory; in particular, in the hot setup (first column), results seem consistent with the sum of the predictions, while in the beam setup (second column), the growth rates seem closer to the resonant scenario, even showing a sharp drop off below $kr_{g,cr}$.

The linear theory also makes claims about the handedness of the unstable modes. 
The helicity of a wave is related to how the wave curls around the background magnetic field line in the direction of ${\bf B}$. Helicity is frame-invariant under Lorentz transformations and is referred to as left- or right-handed depending on how the wave spirals around the magnetic field. 
Helicity can be calculated via the Stokes parameters: $\chi \equiv \sin^{-1}(V/I)/2$, where $I = |\widetilde{B}_\perp|^2$ and $V = -2{\rm Im}[\widetilde{B_y}\widetilde{B_z}^*]$. 
$\chi$ is calculated in degrees and ranges between $\pm 45^{\circ}$, where the positive sign corresponds to right-handed waves, and the negative to left-handed. 
Note that we consider helicity rather than polarization, because determining polarization requires both a specified frame of reference and information about the speed of the waves, which presents an issue for the (purely growing) non-resonant modes.

\begin{figure}[htb]\centering
\includegraphics[trim=1px 0px 0px 1px, clip=true, width=.48\textwidth]{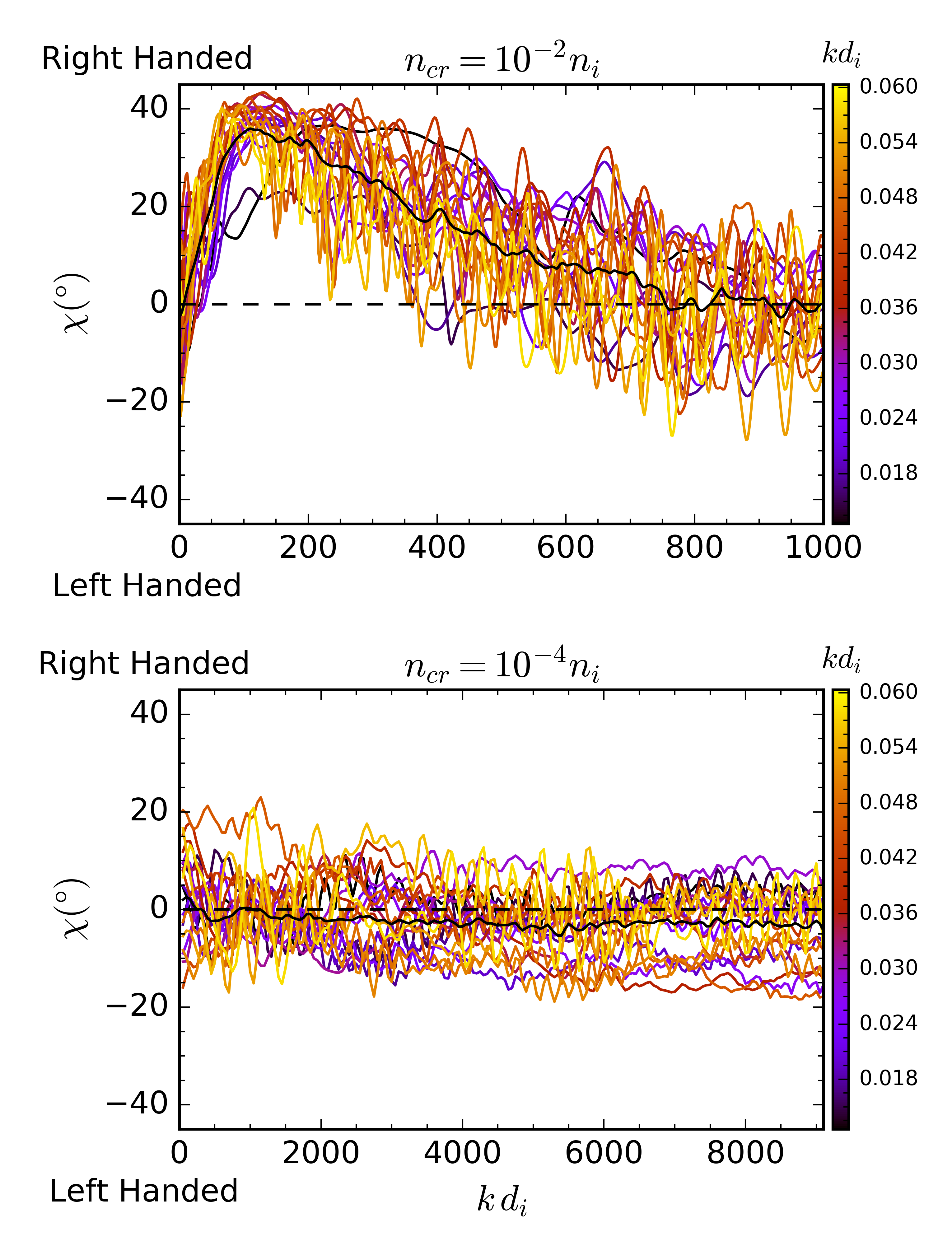}
\includegraphics[trim=1px 0px 0px 1px, clip=true, width=.48\textwidth]{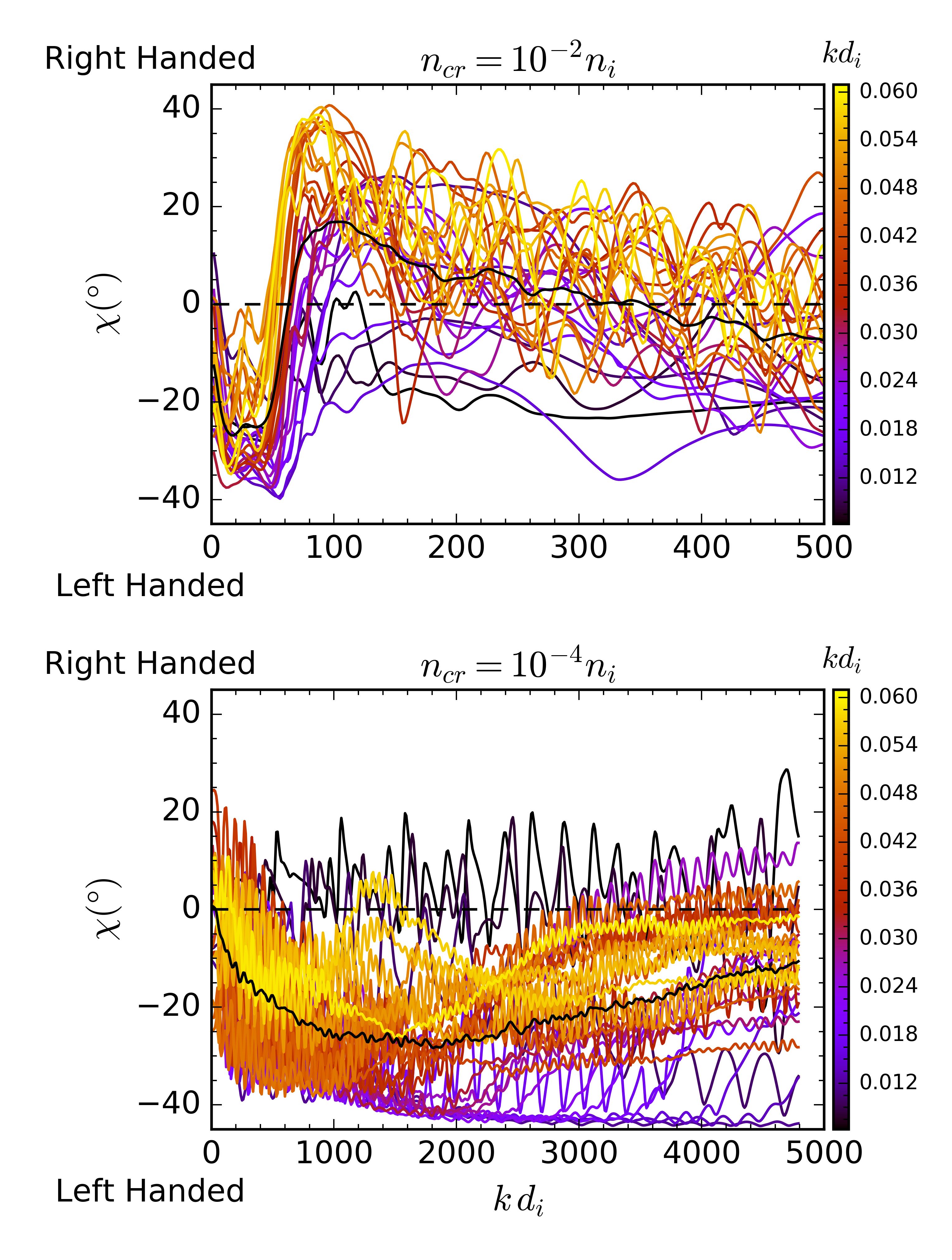}
\caption{{\small Magnetic helicity ($\chi$) for hot/beam (left/right) CR distributions and non-resonant/resonant (top/bottom) simulations as a function of $k$ and time. The black line shows the $k-$averaged helicity.}}\label{fig:pol}
\end{figure}

The helicity for four of the simulations ($n_{cr} = 10^{-2}, 10^{-4}$, for both setups) is shown in Fig.~\ref{fig:pol}. 
Each line is plotted in time and corresponds to a different value of $k$ as indicated by the color bar. The solid black line is the helicity averaged over all the shown wavenumbers.
In the hot setup (left column), there are two distinctly different helicities that develop in time. 
In the higher density CR run (upper panel), we find that within the first $100 \Omega_{ci}^{-1}$ the fluctuations that develop are clearly right-handed for a wide range of $k$ values, as $\chi \sim 40^\circ$. 
This is consistent with the non-resonant instability, which is  driven by the return current in the background electrons, so that unstable modes are resonant with the current carriers.
The bottom panel shows the helicity as a function of time for the weak-current simulation ($n_{cr} = 10^{-4}n_i$), where $\chi \sim 0$ for the entire simulation. 
This is consistent with the fact that both the right-handed and left-handed modes are growing at the same rate.
Physically speaking, this occurs because CRs are streaming in both the positive and negative $x$ directions. 
The small deviation from zero in the left-handed direction is likley related to the net drift velocity ($v_D/v_{cr} \sim v_D/c \sim 0.1$), which gives CRs a small preference in the streaming direction. 
In typical systems $c/v_A\gg1$, hence this asymmetry should be negligible.

In the beam configuration (right column of Fig.~\ref{fig:pol}), the left/right symmetry of the resonant case ($n_{cr} = 10^{-4} n_i$) is no longer present. 
Over the first $1,000 \Omega_{ci}^{-1}$, we find that modes with $k r_{g,cr} > 1$, or equivalently $kd_i > 10^{-2}$, grow with a preferential left handedness, while modes with $k r_{g,cr} < 1$ (dark purple lines) have no preferred helicity.
Since the initial CR distribution contains particles with $\mu>0$, only left-handed waves will resonate with the CRs; at later times $\chi\to 0$, though. 
This is likely a consequence of  the deformation of the initial beam distribution: the resonant waves pitch-angle scatter the beam, reducing the drift velocity and allowing CRs to travel in the $-x$ direction and amplify right-handed modes. 
This explanation is consistent with the fact that the magnetic energy continues to increase with a slightly different growth rate, as indicated by the black line around $2,000 \Omega_{ci}^{-1}$ in the right panel of Fig.~\ref{fig:B}.

Finally, the upper right panel in Fig.~\ref{fig:pol} corresponds to the beam setup for a larger CR density ($n_{cr} = 10^{-2} n_i$) and shows the most notable temporal evolution of helicity of all of the simulations. 
Very early in the simulation ($< 20 \Omega_{ci}^{-1}$), left-handed waves, consistent with the resonant instability, grow. 
Then, slightly later  ($> 30 \Omega_{ci}^{-1}$), right-handed modes, consistent with the non-resonant instability, grow and overtake the left-handed modes. 
We see a small dependence on $k$ for the growth of the non-resonant instability. 
Large values of $kr_{g,cr} \gtrapprox 2$ have $\chi \sim +40^\circ$, while modes closer to the resonant mode have $\chi \sim 0^\circ$. This complex structure can be understood by examining the corresponding $\omega-k$ diagram in the upper right panel in Fig.~\ref{fig:growth}.
Recalling that the red and blue dashed lines are the respective non-resonant and resonant theory predictions, it is clear that both instabilities have comparable maximum growth rates, but in different parts of $k$ space. 
The longer wavelength modes grow very early in the simulation (from $0 \sim 20 \Omega_{ci}^{-1}$), likely leading to the disruption of the CR distribution function, which allows for the non-resonant instability to become dominant at later times (from $20 \sim 60 \Omega_{ci}^{-1}$), until the instability saturates completely and the growth stops ($> 60$ $\Omega_{ci}^{-1}$). 
Note that the growth rates for this run were calculated between $20 \text{ to } 60 \Omega_{ci}^{-1}$, which is why the observed growth rates agree with the non-resonant prediction in the upper right panel of Fig.~\ref{fig:growth}. 
If the calculation window were centered around earlier times, the growth rates would agree with the resonant prediction.

\section{Conclusion}
We used hybrid simulations preformed with \dHybridR~to test the linear theory of several different CR streaming instabilities. 
We tested two different CR distribution functions; a hot power law distribution with a $p^{-4.5}$ slope drifting relative to the background population and a beam-like, nearly mono-energetic distribution. 
We calculated growth rates and helicity of the generated magnetic fluctuations for each CR configuration, changing the CR density relative to the background ions to transition from the non-resonant (or Bell) to resonant streaming inabilities. 
We find that for an appropriate period in the growth of these fluctuations, the growth and helicity of these modes agrees well with linear theory.

In forthcoming works we will also characterize the saturation of the instabilities, the rate at which the background plasma is heated by wave damping, and the possibility of generating CR-driven winds and breezes.

\bibliographystyle{JHEP}
{\footnotesize \bibliography{Total}}

\end{document}